\documentclass[aps,pre,reprint,showkeys,groupedaddress,amsmath,amssymb,floatfix]{revtex4-2}

\usepackage{graphicx}
\usepackage{booktabs}
\usepackage{amsthm}
\usepackage{amsfonts}
\usepackage{mathrsfs}
\usepackage{multirow}
\usepackage{array}
\usepackage{capt-of}

\newtheorem{theorem}{Theorem}
\newtheorem{proposition}[theorem]{Proposition}

\begin{document}
\title{Spatial constraints and multi-peak epidemic dynamics in random geometric metapopulation networks}

\author{Guillermo Capobianco}
\email{guillermo.capobianco@gmail.com}
\affiliation{Instituto de Matem\'atica (INMABB), Departamento de Matem\'atica, Universidad Nacional del Sur (UNS)--CONICET, Alem 1253, Bah\'ia Blanca, 8000 Buenos Aires, Argentina}
\author{Romina Cobiaga}
\thanks{These authors contributed equally to this work.}
\affiliation{Instituto de Matem\'atica (INMABB), Departamento de Matem\'atica, Universidad Nacional del Sur (UNS)--CONICET, Alem 1253, Bah\'ia Blanca, 8000 Buenos Aires, Argentina}
\author{Beatriz Marr\'on}
\thanks{These authors contributed equally to this work.}
\affiliation{Instituto de Matem\'atica (INMABB), Departamento de Matem\'atica, Universidad Nacional del Sur (UNS)--CONICET, Alem 1253, Bah\'ia Blanca, 8000 Buenos Aires, Argentina}
\author{Walter Reartes}
\thanks{These authors contributed equally to this work.}
\affiliation{Instituto de Matem\'atica (INMABB), Departamento de Matem\'atica, Universidad Nacional del Sur (UNS)--CONICET, Alem 1253, Bah\'ia Blanca, 8000 Buenos Aires, Argentina}

\begin{abstract}
Spatial structure can strongly affect epidemic propagation, but its role in shaping aggregate epidemic curves is often obscured when metapopulation models are studied on non-spatial or well-mixed network structures.
Here, we investigate whether spatially constrained connectivity can generate multi-peak aggregate epidemic dynamics that are strongly reduced or absent in non-spatial networks with comparable connectivity.
We simulate SIR metapopulation dynamics on random geometric graphs (RGGs) and compare their epidemic trajectories with those generated by degree-matched Erd\H{o}s--R\'enyi, Watts--Strogatz, and Barab\'asi--Albert networks.
We further consider toroidal RGGs to distinguish the effects of local geometric connectivity from those associated with hard boundaries, and examine two alternative formulations of inter-node transmission as a robustness check.
Across network ensembles, RGGs produce delayed spatial invasion and more temporally staggered local epidemic propagation than degree-matched non-spatial networks.
These differences in the timing of local outbreaks can generate broad, shouldered, or multi-peak aggregate infected trajectories.
The trajectory and spatiotemporal comparisons consistently show a more localized and asynchronous pattern of propagation in spatially constrained networks.
Toroidal RGG controls show that multi-peak behavior persists under periodic boundary conditions, although it is reduced relative to square RGGs, indicating that hard boundaries amplify the phenomenon but do not fully account for it.
Together, these results support the interpretation that local geometric connectivity can transform simple local epidemic dynamics into complex aggregate temporal patterns through delayed and asynchronous spatial propagation.
Thus, the spatial organization of connections, rather than connectivity alone, can provide a structural mechanism for generating multi-peak epidemic dynamics in metapopulation networks.
\end{abstract}


\maketitle
\section{Introduction}\label{sec1}

Understanding how spatial structure shapes epidemic propagation remains a central problem in mathematical epidemiology.
Aggregate epidemic curves often display broad peaks, shoulders, plateaus, or multiple epidemic waves, even when the local disease dynamics are relatively simple.
Such temporal complexity is commonly interpreted in terms of behavioral changes, parameter heterogeneity, seasonality, or external forcing.
However, in spatially structured populations, part of this complexity may also arise from the geometry of transmission pathways: local outbreaks need not occur synchronously across regions, and the aggregate signal can reflect the superposition of temporally staggered local epidemics.

Metapopulation epidemic models provide a natural framework for studying this problem.
In these models, each node represents a subpopulation, such as a city, neighborhood, or region, and local epidemic dynamics are coupled through mobility, contact, or effective transmission processes between nodes \cite{ColizzaVespignani2008,Poletto2013}.
A central lesson from this literature is that epidemic spread in a metapopulation is not determined only by the local reproduction number: global invasion also depends on the network of subpopulations, mobility rates, and topological heterogeneity \cite{ColizzaVespignani2008}.
Thus, network structure and inter-population coupling are part of the epidemic mechanism, rather than secondary modeling details.

Classical compartmental models such as the SIR model assume homogeneous mixing within a population, whereas network-based metapopulation models allow the interaction structure between subpopulations to be represented explicitly \cite{pastorsatorras2015,kiss23,bonaccorsi15,selleyfani15,wang08,knipl13,keeling2011networks}.
Data-driven mobility models have further shown that the spatial distribution of hosts and their movement patterns are essential for understanding epidemic spread across regions.
In such models, subpopulations are localized in space and connected by mobility flows that may represent commuting, travel, recurrent movement, or time spent at destinations \cite{Poletto2013,cosner2009}.

The mathematical representation of mobility itself can also substantially
affect epidemic dynamics. Citron et al.~\cite{Citron2021} explicitly compared
alternative formulations of human movement within metapopulation epidemic
models and showed that different movement assumptions can lead to markedly
different epidemiological outcomes. More broadly, Lu et al.~\cite{Lu2026}
review recent approaches for incorporating human mobility into epidemic
models, highlighting the diversity of mobility representations and their
role in shaping spatiotemporal disease dynamics. Related multi-patch
frameworks for vector-borne diseases have also incorporated both host
commuting and vector migration, using complementary Lagrangian and Eulerian
descriptions of movement \cite{gao2024}.

These approaches are crucial for realistic forecasting and for modeling specific systems.
However, empirical mobility networks often combine many structural features at once, including long-range travel, hub structure, heterogeneous fluxes, recurrent mobility, and population imbalance.
This can make it difficult to isolate how a single structural mechanism affects epidemic dynamics.

Controlled spatial network ensembles provide a complementary approach.
In this work, we focus on random geometric graphs (RGGs) as controlled spatial network ensembles.
RGGs are not introduced here as empirical reconstructions of a particular transportation, contact, or mobility system.
Instead, they are used as mathematically well-defined spatial networks that isolate the effect of local geometric connectivity on epidemic propagation.
In an RGG, nodes are embedded in a metric space and edges are formed according to spatial proximity, typically when the distance between two nodes is below a fixed connection radius \cite{dall2002,penrose2003,estrada2016}.
This construction suppresses long-range links and forces transmission to
proceed through local graph neighborhoods, generating spatial correlations,
high clustering, and more spatially localized propagation patterns than
many non-geometric networks with comparable mean degree.

Spatial epidemic modeling has emphasized the importance of spatial structure and the challenge of constructing network models that capture local correlations, clustering, and spatial propagation while remaining analytically and computationally tractable \cite{riley2015,pellis2015}.
In particular, spatial network models can generate organized propagation patterns, including traveling-wave-like solutions, showing that spatial structure can affect not only final epidemic size or threshold behavior, but also the temporal organization of epidemic spread \cite{lang2018}.
This perspective motivates the use of RGGs as controlled spatial ensembles for studying how local geometric constraints shape invasion timing, local outbreak synchronization, and aggregate epidemic trajectories.

Most network comparisons in epidemic modeling emphasize quantities such as epidemic thresholds, final outbreak size, or the effect of degree heterogeneity.
Here, instead, we focus on the temporal and spatiotemporal organization of the epidemic trajectory.
The central question of this work is whether spatially constrained connectivity can delay epidemic invasion, desynchronize local outbreaks, generate sustained spatiotemporal propagation, and produce broad, shouldered, or multi-peak aggregate epidemic curves.
This question is motivated by the fact that complex aggregate incidence profiles can arise even when local epidemic dynamics are simple, suggesting that the geometry of transmission pathways may itself be a mechanism for producing temporal complexity.

To address this question, we study SIR metapopulation dynamics on square RGGs and compare them with degree-matched Erd\H{o}s--R\'enyi, Watts--Strogatz, and Barab\'asi--Albert network ensembles.
The degree-matched baselines allow us to compare networks with broadly similar average connectivity while changing the organization of links.
We also introduce toroidal RGGs as a periodic-boundary control.
This control removes hard boundaries while preserving local geometric connectivity, allowing us to distinguish boundary amplification from the more general effect of spatially constrained paths.

We consider two formulations of inter-node coupling.
In the mobility-consistent mixing model (MCM), individuals move between connected subpopulations and mix at their destination before transmission occurs.
In the effective-coupling model (EC), infection pressure is transmitted across network edges without explicitly representing population movement.
These two formulations are not treated as competing empirical reconstructions of a specific mobility system.
Rather, they are used as mechanistic alternatives to test whether the dynamical signatures induced by spatial network structure are robust to how inter-population transmission is represented.
Thus, the primary comparison in this paper is topological---spatially constrained RGGs versus degree-matched baseline networks---while the EC/MCM comparison plays a secondary role as a robustness check across coupling mechanisms.

Our main finding is that spatially constrained RGGs produce delayed and more
temporally staggered epidemic propagation than degree-matched non-geometric
baselines. When local outbreaks become sufficiently desynchronized, their
aggregate contribution can generate prolonged, shouldered, or multi-peak
epidemic trajectories. This behavior is most pronounced in square RGGs,
is reduced but persists in toroidal RGGs, and is absent from all degree-matched non-geometric baselines across the explored parameter sweep.

The contributions of this work are therefore threefold.
First, we provide a controlled ensemble comparison showing how RGG-like
spatial constraints affect aggregate epidemic trajectories relative to
degree-matched network baselines.
Second, we use toroidal RGGs as a periodic-boundary control to distinguish
the contribution of hard boundaries from the broader effect of local
geometric connectivity.
Third, we show that the qualitative RGG-driven behavior persists under two
different coupling formulations, although its magnitude depends on how
inter-node transmission is represented.
Together, these results support local geometric connectivity as a structural
mechanism linking spatial network organization to delayed propagation,
local outbreak asynchrony, and complex aggregate epidemic time series.

\section{Model and methods}\label{sec:model}

We consider a metapopulation SIR model on a fixed, undirected network with $n$ nodes.
Node \(i\) represents a subpopulation of constant size \(N_i>0\), partitioned into susceptible, infected, and removed compartments \(S_i(t),I_i(t),R_i(t)\), with \(S_i(t)+I_i(t)+R_i(t)=N_i\). We neglect births, deaths unrelated to the disease, and permanent migration. Inter-node interactions are mediated by a coupling mechanism that specifies how susceptibles in one node are exposed to infectives from other nodes.

\subsection{Mobility-consistent mixing (MCM) formulation}\label{subsec:new}
To model short-term daily movements (e.g., commuting or visits), we introduce a mobility (mixing) matrix
\(P=(P_{ij})\), where \(P_{ij}\in[0,1]\) denotes the fraction of time (or fraction of individuals) from home node \(i\) that is physically present in destination node \(j\) during the relevant mixing timescale. We assume that \(P\) is row-stochastic,
\begin{equation}\label{eq:row_stochastic}
\sum_{j=1}^n P_{ij}=1,\qquad P_{ij}\ge 0,
\end{equation}
and, to preserve the underlying topology, that mobility occurs only along edges of the network, %
i.e., \(P_{ij}=0\) if \(i\) and \(j\) are not adjacent and \(i\neq j\).
This formulation is closely related to residence-time or Lagrangian mobility models, in which individuals retain a home population while spending fractions of time at different destinations \cite{cosner2009}. %

\paragraph{Population present at each destination.}
Under this mobility mechanism, the total population physically present at node \(j\) is
\begin{equation}\label{eq:Npres}
N_j^{\mathrm{pres}}=\sum_{\ell=1}^n P_{\ell j}N_\ell,
\end{equation}
which is constant in time under our assumptions. Similarly, the total number of infected individuals physically present at node \(j\) is
\begin{equation}\label{eq:Ipres}
I_j^{\mathrm{pres}}(t)=\sum_{\ell=1}^n P_{\ell j}I_\ell(t).
\end{equation}
We assume homogeneous mixing within each destination \(j\) during the mixing period, so that susceptibles present in \(j\) are exposed to the entire infectious pool \(I_j^{\mathrm{pres}}(t)\).

\paragraph{Force of infection and equations.}
A susceptible individual from home node \(i\) spends a fraction \(P_{ij}\) of time in destination \(j\), and while there experiences per-capita infection hazard proportional to the infected fraction present at \(j\), \(I_j^{\mathrm{pres}}/N_j^{\mathrm{pres}}\). Therefore, the effective force of infection for home node \(i\) is
\begin{equation}\label{eq:lambda_new}
\lambda_i(t)=\beta\sum_{j=1}^n P_{ij}\,\frac{I_j^{\mathrm{pres}}(t)}{N_j^{\mathrm{pres}}}.
\end{equation}
The metapopulation SIR dynamics are then
\begin{equation}\label{eq:SIR_new_lambda}
\begin{aligned}
\frac{dS_i}{dt}&=-\lambda_i(t)\,S_i,\\
\frac{dI_i}{dt}&=\lambda_i(t)\,S_i-\gamma I_i,\\
\frac{dR_i}{dt}&=\gamma I_i, \qquad i=1,\dots,n.
\end{aligned}
\end{equation}

\paragraph{Connectivity-matrix form.}
Substituting \eqref{eq:Ipres} into \eqref{eq:lambda_new} yields
\[
\lambda_i(t)=\beta\sum_{\ell=1}^n c_{i\ell}\,I_\ell(t),
\qquad
c_{i\ell}=\sum_{j=1}^n \frac{P_{ij}P_{\ell j}}{N_j^{\mathrm{pres}}},
\]
so that \eqref{eq:SIR_new_lambda} can be written compactly as
\begin{equation}\label{eq:SIR_new_C}
\begin{aligned}
\frac{dS_i}{dt}&=-\beta S_i\sum_{\ell=1}^n c_{i\ell}I_\ell,\\
\frac{dI_i}{dt}&=\beta S_i\sum_{\ell=1}^n c_{i\ell}I_\ell-\gamma I_i,\\
\frac{dR_i}{dt}&=\gamma I_i.
\end{aligned}
\end{equation}
with \(C=(c_{i\ell})\) fully determined by the mobility matrix \(P\).

\paragraph{Choice of \(P\) on a graph.}
A simple local-mobility rule, parameterized by \(\delta\in[0,1]\), is: an individual stays at home with probability \(1-\delta\) and visits one of its neighbors uniformly at random with probability \(\delta\). Let \(A\) be the adjacency matrix and \(k_i=\sum_j A_{ij}\) the degree of node \(i\). Then
\begin{equation}\label{eq:P_local}
P_{ij}=
\begin{cases}
1-\delta, & j=i,\\[2pt]
\delta/k_i, & A_{ij}=1,\\[2pt]
0, & \text{otherwise.}
\end{cases}
\end{equation}
This choice ensures \eqref{eq:row_stochastic} and restricts movement to adjacent nodes.

\subsubsection{Basic reproduction number under mobility-consistent mixing}\label{subsec:R0_new}
We define the basic reproduction number \(R_0\) via the next-generation matrix method. In the present framework, linearizing \eqref{eq:SIR_new_C} around the disease-free equilibrium \(S_i=N_i\) yields a next-generation matrix of the form \(K=(\beta/\gamma)\,\mathrm{diag}(N)\,C\), where \(C\) is induced by \(P\). The mobility-consistent formulation yields a topology-independent threshold:

\begin{proposition}[Mobility-consistent mixing yields $R_0=\beta/\gamma$]
Assume $P$ is row-stochastic $\left(\sum_{j}P_{ij}=1\right)$ and $N_i>0$ for all $i$.
Let
\[
c_{i\ell}=\sum_{j=1}^n \frac{P_{ij}P_{\ell j}}{N_j^{\mathrm{pres}}},
\qquad
N_j^{\mathrm{pres}}=\sum_{\ell=1}^n P_{\ell j}N_\ell,
\]
and define the next-generation matrix
\[
K=\frac{\beta}{\gamma}\,\mathrm{diag}(N)\,C,
\qquad C=(c_{i\ell}).
\]
Then $\rho(\mathrm{diag}(N)C)=1$ and consequently $R_0=\rho(K)=\beta/\gamma$.
\end{proposition}

\begin{proof}
Linearizing the infected equations at the disease-free equilibrium $S_i=N_i$ gives
\[
\frac{dI_i}{dt}\approx \beta N_i\sum_{\ell=1}^n c_{i\ell}I_\ell-\gamma I_i,
\]
so $K=(\beta/\gamma)\mathrm{diag}(N)C$.

Consider $B:=C\,\mathrm{diag}(N)$. Its row sums satisfy
\begin{align*}
\sum_{\ell=1}^n B_{i\ell}
&=\sum_{\ell=1}^n c_{i\ell}N_\ell\\
&=\sum_{\ell=1}^n\sum_{j=1}^n
\frac{P_{ij}P_{\ell j}}{N_j^{\mathrm{pres}}}N_\ell\\
&=\sum_{j=1}^n \frac{P_{ij}}{N_j^{\mathrm{pres}}}
\sum_{\ell=1}^n P_{\ell j}N_\ell\\
&=\sum_{j=1}^n P_{ij}=1.
\end{align*}
hence $B$ is row-stochastic. Therefore $1$ is an eigenvalue of $B$ (with eigenvector
$\mathbf{1}$) and $\rho(B)\le \|B\|_\infty=1$, implying $\rho(B)=1$.
Since $\mathrm{diag}(N)C$ and $C\,\mathrm{diag}(N)$ have the same nonzero eigenvalues,
$\rho(\mathrm{diag}(N)C)=\rho(B)=1$. Thus
\[
R_0=\rho(K)=\frac{\beta}{\gamma}\rho(\mathrm{diag}(N)C)=\frac{\beta}{\gamma}.
\]
\end{proof}

\noindent
This result highlights a key conceptual point: under mobility-consistent mixing and homogeneous node sizes, spatial structure redistributes exposure through \(P\) and the induced operator \(C\), but does not amplify the linearized epidemic threshold. Network effects can still manifest in transient dynamics and synchronization patterns, which we investigate numerically.

\subsection{Effective-coupling (EC) formulation}\label{subsec:old}
We next introduce a symmetric effective-coupling (EC) formulation, which provides a conceptually distinct description of transmission across neighboring subpopulations. Rather than modeling short-term mobility and mixing at shared destinations, this approach represents infection pressure through direct cross-node interactions.
The force of infection acting on node \(i\) is taken as
\begin{equation}\label{eq:lambda_old}
\lambda_i(t)=\beta\sum_{j=1}^n c_{ij}^{\mathrm{\scriptscriptstyle EC}}\,I_j(t),
\end{equation}
with a symmetric connectivity matrix
\begin{equation}\label{eq:C_old}
\begin{aligned}
c_{ij}^{\mathrm{\scriptscriptstyle EC}}
&=\frac{\big(\delta^K_{ij}+\delta A_{ij}\big)\,(N_i+N_j)}{2N_iN_j}\\
&=\frac12\big(\delta^K_{ij}+\delta A_{ij}\big)
\left(\frac{1}{N_i}+\frac{1}{N_j}\right).
\end{aligned}
\end{equation}
where \(A\) is the adjacency matrix of the network, \(\delta^K_{ij}\) is the Kronecker delta, and \(\delta\ge 0\) controls the strength of cross-node coupling.

Equation \ref{eq:C_old} is introduced here as a phenomenological effective-coupling
construction to represent symmetric edge-mediated infection pressure between
neighboring subpopulations. The population-size factor
$\displaystyle\frac{N_i+N_j}{2N_iN_j}$
provides a symmetric normalization of this interaction and can be written as
the reciprocal of the harmonic mean of \(N_i\) and \(N_j\).

Harmonic-mean-type constructions have previously been considered in nonlinear
epidemic incidence formulations
\cite{Rahman2019,KhanZarin2020,Chawla2025}.
These formulations use the harmonic mean to define nonlinear incidence terms
involving interacting epidemiological compartments or populations. In the
present EC formulation, by contrast, the reciprocal harmonic-mean factor acts
as a population-size normalization of the effective cross-node coupling.

The corresponding metapopulation dynamics read
\begin{equation}\label{eq:SIR_old_C}
\begin{aligned}
\frac{dS_i}{dt}&=-\beta S_i\sum_{j=1}^n c_{ij}^{\mathrm{\scriptscriptstyle EC}}I_j,\\
\frac{dI_i}{dt}&=\beta S_i\sum_{j=1}^n c_{ij}^{\mathrm{\scriptscriptstyle EC}}I_j-\gamma I_i,\\
\frac{dR_i}{dt}&=\gamma I_i.
\end{aligned}
\end{equation}

\subsubsection{Basic reproduction number under EC}\label{subsec:R0_old}
\begin{figure*}[t]
  \centering
  \includegraphics[width=0.69\textwidth]{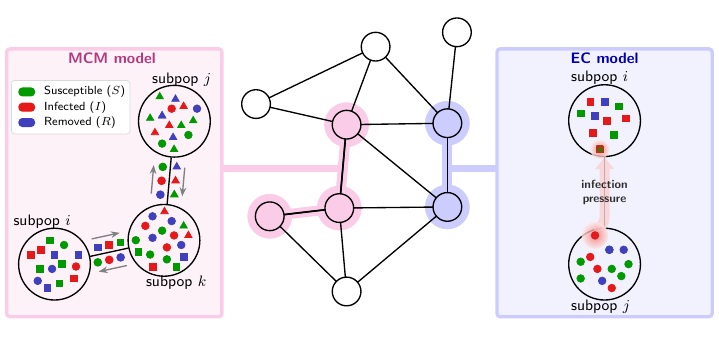}
  \caption{ Schematic representation of the two coupling formulations used in the metapopulation model. In the mobility-consistent mixing model (MCM), individuals from neighboring subpopulations move and mix at their destination nodes before transmission occurs. Thus, infection risk depends on the composition of the population present at each destination. In the effective-coupling model (EC), no explicit movement of individuals is represented; instead, infected individuals in one subpopulation exert an effective infection pressure on susceptible individuals in adjacent subpopulations. Colors denote epidemiological states: susceptible (\(S\)), infected (\(I\)), and removed (\(R\)). Marker shapes in the MCM panel indicate the home subpopulation of each individual: squares, triangles, and circles correspond to home subpopulations \(i\), \(j\), and \(k\), respectively. Arrows in the MCM panel indicate movement between subpopulations, whereas the arrow in the EC panel represents effective infection pressure rather than movement.}
  \label{fig:coupling_schematic}
\end{figure*}

Linearizing \eqref{eq:SIR_old_C} at the disease-free equilibrium \(S_i=N_i\) yields the next-generation matrix
\begin{equation}\label{eq:K_old}
K_{\mathrm{\scriptscriptstyle EC}}=\frac{\beta}{\gamma}\,\mathrm{diag}(N)\,C_{\mathrm{\scriptscriptstyle EC}},
\qquad C_{\mathrm{\scriptscriptstyle EC}}=\big(c_{ij}^{\mathrm{\scriptscriptstyle EC}}\big),
\end{equation}
and hence
\begin{equation}\label{eq:R0_old_general}
R_0^{\mathrm{\scriptscriptstyle EC}}=\rho\!\left(K_{\mathrm{\scriptscriptstyle EC}}\right)
=\frac{\beta}{\gamma}\,\rho\!\left(\mathrm{diag}(N)\,C_{\mathrm{\scriptscriptstyle EC}}\right).
\end{equation}
In the homogeneous-population case \(N_i\equiv N\), \eqref{eq:C_old} reduces to
\begin{equation}\label{eq:C_old_hom}
C_{\mathrm{\scriptscriptstyle EC}}=\frac{1}{N}\big(\mathbb{I}+\delta A\big),
\end{equation}
so that
\begin{equation}\label{eq:R0_old_hom}
R_0^{\mathrm{\scriptscriptstyle EC}}=\frac{\beta}{\gamma}\,\rho(\mathbb{I}+\delta A)
=\frac{\beta}{\gamma}\big(1+\delta\lambda_{\max}(A)\big),
\end{equation}
where \(\lambda_{\max}(A)\) is the largest eigenvalue of \(A\). Thus, in the effective-coupling formulation the epidemic threshold depends explicitly on the adjacency spectrum, illustrating how different coupling assumptions can induce markedly different topology dependence in \(R_0\) and related outcomes.
Figure~\ref{fig:coupling_schematic} provides a schematic comparison of the
two coupling formulations, highlighting the explicit movement and
destination-based mixing represented in MCM and the edge-mediated infection
pressure represented in EC.

\paragraph{Relation between the two couplings.}

The \textsc{EC} model captures cross-subpopulation transmission along network edges through a symmetric, pairwise interaction term with population-size normalization. Rather than representing short-term commuting to shared destinations, it models direct interaction pressure between neighboring subpopulations. In this interpretation, transmission arises from localized cross-interface contacts, for instance through social, commercial, or environmental proximity across adjacent regions. The resulting coupling can therefore be viewed either as a phenomenological effective description of cross-infection pressure between neighbors, or as a coarse-grained representation of spatially localized transmission processes (e.g., vector-mediated or environmentally mediated contagion) acting at the interface between populations.

By contrast, the \textsc{MCM} formulation derives transmission from an explicit mobility matrix \(P\) satisfying \(\sum_j P_{ij}=1\), corresponding to short-term commuting and mixing at common destinations. Infection occurs at activity centers where individuals from multiple origins interact, and the effective operator \(C\) reflects this destination-based mixing while preserving population fractions.

In this sense, the two couplings represent complementary modeling philosophies: mobility-driven mixing versus localized cross-interface transmission (Table~\ref{tab:MCMEC}).

\begin{table}[t]
\caption{\label{tab:MCMEC}
Conceptual comparison of the MCM and EC coupling formulations.}
\centering
\footnotesize
\begin{ruledtabular}
\begin{tabular}{lll}
Feature & \textsc{MCM} & \textsc{EC} \\
\hline
Transmission & Activity centers & Neighbor interfaces \\
Mobility & Explicit & Effective local coupling \\
$R_0$ & $\beta/\gamma$ & Topology dependent \\
Variability & Mobility matrix $P$ & Adjacency matrix $A$ \\
\end{tabular}
\end{ruledtabular}
\end{table}

\subsection{Random geometric graphs (RGG)}\label{subsec:rgg}
A common approach to incorporate spatial effects is to discretize a region as a lattice; alternatively, random geometric graphs provide a flexible framework for spatially constrained yet heterogeneous connectivity \cite{dall2002,penrose2003}. An RGG is constructed by sampling \(n\) node positions independently and uniformly in the unit square \([0,1]^2\), and connecting two nodes if their Euclidean distance is below a prescribed connectivity radius \(r\). In our simulations, \(r\) is the main geometric control parameter: increasing \(r\) increases connectivity and reduces spatial fragmentation.

RGGs naturally encode locality, high clustering, and degree--degree correlations induced by spatial density fluctuations, features that are absent or much weaker in classical non-spatial random graph models \cite{badham10,newman2002,penrose2003}. These structural properties have been shown to affect epidemic behavior and propagation pathways \cite{preciado2016,preciado2009spectral,sazonov2011travelling}. In particular, degree assortativity in two-dimensional RGGs exhibits intrinsic positive correlations in the large-\(n\) limit \cite{antonioni2012}, reflecting the fact that nodes in denser regions tend to connect to neighbors that are also in dense regions.

Unlike Erd\H{o}s--R\'enyi or Barab\'asi--Albert models, RGGs inherently incorporate spatial constraints and typically exhibit high clustering and nontrivial degree correlations, which makes them a useful testbed for geographically localized epidemic interactions \cite{badham10,newman2002,penrose2003}. In particular, in Erd\H{o}s--R\'enyi graphs the assortativity coefficient is typically close to zero, indicating weak degree correlation between adjacent nodes, whereas spatial embedding in RGGs can induce systematic degree--degree correlations through local density fluctuations.

\paragraph{Connectivity and connectedness.}
Because epidemic propagation is constrained by connectivity, we generate RGG realizations under the additional requirement that the graph be connected. For each parameter pair \((n,r)\), we use rejection sampling (regenerating node positions) until a connected realization is obtained. This ensures that observed epidemic outcomes are not trivially limited by disconnected components.

\subsection{Network ensembles and degree-matched baselines}
\label{subsec:network_ensembles}

We considered five network topologies: square random geometric graphs (RGGs), toroidal RGGs, Erd\H{o}s--R\'enyi (ER), Watts--Strogatz (WS), and Barab\'asi--Albert (BA) networks \cite{penrose2003,gilbert1959,erdos1959,watts1998,barabasi1999}.
Square RGGs were generated by placing \(n\) nodes uniformly at random in the unit square and connecting two nodes when their Euclidean distance was smaller than a prescribed radius \(r\).
RGGs are used here as controlled spatial network ensembles, not as empirical reconstructions of a particular mobility system.

To control for boundary effects while preserving local geometric connectivity,
we also generated toroidal RGGs with periodic boundary conditions on
\([0,1)^2\). In the toroidal case, distances were computed periodically.
For each square-RGG realization, its realized mean degree was used as the
target mean degree for the corresponding toroidal RGG. If
\( k_{\mathrm{target}}\) denotes this target value, we used
\[
r_{\mathrm{torus}}
=
\sqrt{\frac{ k_{\mathrm{target}}}{\pi(n-1)}}.
\]
Thus, each toroidal RGG was degree-matched in expectation to the
corresponding square-RGG realization, and consequently the toroidal ensemble
was degree-matched in expectation to the square-RGG ensemble.

As non-spatial degree-matched controls, we generated ER, WS, and BA networks.
For each square RGG configuration, the observed mean degrees of the RGG realizations were used to define target mean degrees for the baseline ensembles.
Specifically, the RGG mean degrees were grouped into bins of width \(\Delta k=2.0\), and only bins containing at least five RGG realizations were retained.
For each retained bin, we generated independent ER, WS, and BA networks with target mean degree \(k_{\mathrm{target}}\) equal to the bin-averaged RGG mean degree.
ER graphs used $p=\frac{k_{\mathrm{target}}}{n-1},$
WS graphs used an integer nearest-neighbor degree \(k\simeq k_{\mathrm{target}}\), adjusted to be even, with rewiring probability \(p_{\mathrm{rewire}}=0.05\), and BA graphs used
$m\simeq \frac{k_{\mathrm{target}}}{2}.$
This degree-matched design allows us to compare spatial and non-spatial topologies at comparable average connectivity, so that differences in epidemic propagation are not attributable simply to differences in mean degree.

\subsection{Parameter sweep and simulation protocol}
\label{subsec:simulation_protocol}

The parameter sweep used $n\in\{200,400\},$ with square RGG radii
$r\in\{0.095,0.110,0.125\}\quad \text{for }n=200,$
and $r\in\{0.075,0.090,0.105\}\quad \text{for }n=400.$
The epidemiological and coupling parameters were varied over
$\beta\in\{0.20,0.25,0.30\},\qquad \gamma\in\{0.08,0.10,0.12\},\qquad \delta\in\{0.03,0.08,0.12\}.$
This gives \(162\) parameter configurations \((n,r,\beta,\gamma,\delta)\).

For each parameter configuration, we generated 40 independent square RGG
realizations and 40 corresponding degree-matched toroidal RGG realizations,
with the toroidal radius determined separately for each realization as
described above. For the degree-matched ER, WS, and BA baselines, 20
independent networks were generated for each retained mean-degree bin, as
described above.

For every network realization, we simulated the epidemic dynamics under both coupling formulations: the effective-coupling model (EC) and the mobility-consistent mixing model (MCM).
Thus, topology-dependent effects were assessed separately within each coupling formulation by comparing square RGGs and toroidal RGGs with degree-matched ER, WS, and BA baselines.
The primary comparison is topological: spatially constrained networks versus degree-matched non-spatial baselines.
The EC/MCM comparison is used as a robustness dimension, allowing us to test whether topology-dependent propagation patterns persist under different assumptions about inter-node transmission.

Unless otherwise stated, each node represented a subpopulation of equal size \(N_i=1000\), so that the total population was \(N_{\mathrm{tot}}=1000 n\).
Epidemics were initiated by introducing one infected individual at a single seed node \(i_0\), with all remaining individuals initially susceptible.
Thus, $I_{i_0}(0)=1,\qquad S_{i_0}(0)=N_{i_0}-1,$ and, for \(j\neq i_0\),  $I_j(0)=0,\qquad S_j(0)=N_j,$ with \(R_j(0)=0\) for all nodes.

The ODE system was integrated using the explicit Runge--Kutta method of order \(5(4)\) implemented in \texttt{scipy.integrate.solve\_ivp} \cite{virtanen2020scipy}.
Simulations were run over \(t\in[0,300]\) and sampled at unit time resolution \(\Delta t=1\).
Relative and absolute tolerances were set to \(10^{-6}\) and \(10^{-9}\), respectively.

\subsection{Multi-peak detection}
\label{subsec:peaks}

For each simulation, we recorded the aggregate infected trajectory
\[
    I_{\mathrm{tot}}(t)=\sum_{i=1}^{n} I_i(t),
\]
which is the main time series used to characterize aggregate epidemic shape.

To identify multi-peak aggregate epidemic trajectories, we applied a fixed signal-processing procedure to \(I_{\mathrm{tot}}(t)\).
First, the aggregate infected curve was smoothed using a Savitzky--Golay polynomial filter \cite{savitzky1964}, with window length \(w=21\) and polynomial order \(p=3\), as implemented in \texttt{scipy.signal.savgol\_filter}
\cite{virtanen2020scipy}.
Peaks were then identified using \texttt{scipy.signal.find\_peaks} \cite{virtanen2020scipy}.

A local maximum of the smoothed aggregate curve was retained as an epidemic peak only if it satisfied two criteria: a minimum prominence
$\pi_{\min}=0.02\,\max_t I_{\mathrm{tot}}(t)$,
and a minimum temporal separation $d_{\min}=14$
days from neighboring detected peaks.
A trajectory was classified as multi-peak when at least two peaks satisfied these criteria.

The same smoothing and peak-detection parameters were used for both EC and MCM dynamics and for all network topologies.
Thus, differences in multi-peak frequency across topologies or coupling formulations were not introduced by topology-dependent post-processing choices.

As a robustness check, we varied the smoothing window (17--25 days),
polynomial order (2--4), minimum peak separation (10--18 days), and
prominence threshold (1.5--2.5\%). Across 81 combinations of these
parameters, 238 of 240 trajectories (99.2\%) retained the same binary
single-peak/multi-peak classification, indicating that the qualitative
results are insensitive to reasonable variations of the detection criteria.

\subsection{Representative trajectory and heatmap visualizations}
\label{subsec:representative_visualizations}

In addition to ensemble-level summaries, we generated representative trajectory and heatmap visualizations to illustrate the mechanisms observed in the simulations.
These figures are not ensemble averages and are not used as the sole basis for quantitative conclusions.

For the representative trajectory panels, selected parameter configurations were taken from the parameter sweep to illustrate regimes with visible RGG-driven delayed invasion or multi-peak behavior.
For each selected configuration, we generated six independent network realizations for each topology and simulated both EC and MCM dynamics on each realization.
Each plotted curve corresponds to one individual simulation and represents the aggregate infected trajectory \(I_{\mathrm{tot}}(t)\).
Solid curves denote EC simulations and dashed curves denote MCM simulations.
The 2+3 panel layout displays square RGG and toroidal RGG on the first row, and ER, WS, and BA baselines on the second row.

Heatmap panels visualize node-level infected counts \(I_i(t)\) at selected times.
All metapopulations have equal population size within each simulation, and local maxima of \(I_i(t)\) are not constrained to be the same across nodes or networks.
Color scales were fixed within each coupling formulation, so colors are comparable across topologies within EC and within MCM.

For the representative trajectory and heatmap visualizations shown in
Figs.~2, 4, and 5, we used \(n=200\) nodes with equal local population
size \(N_i=2500\), and fixed
\(\beta=0.20\), \(\gamma=0.10\), and \(\delta=0.03\).
The square RGG radius was \(r=0.110\), while the corresponding toroidal
radius was determined using the degree-matching procedure described in
Section~\ref{subsec:network_ensembles}.
These simulations were run up to \(T=450\) days with daily output
\(\Delta t=1\).
The resulting network parameters and realized mean degrees are summarized
in Table~2.

\subsection{Implementation and reproducibility}
\label{subsec:implementation}

All simulations and analyses were implemented in Python using NumPy, SciPy, NetworkX, and Matplotlib \cite{virtanen2020scipy,hagberg2008networkx}.
Network generation included connectedness checks, toroidal RGG construction using \texttt{scipy.spatial.cKDTree} with periodic boundary conditions \cite{virtanen2020scipy},
 and degree-matched baseline generation.
The computational pipeline then simulated the metapopulation epidemic model
under both EC and MCM dynamics, applied the fixed peak-detection protocol,
and summarized the resulting simulation outcomes.
The source code and generated datasets are available from the corresponding author upon request.

\section{Results}\label{sec:results}

In this section we summarize the main numerical findings obtained from the network ensembles and simulation protocol described in Sections~\ref{subsec:rgg}--\ref{subsec:peaks}.
All simulations are deterministic conditional on a given network realization.
Therefore, variability across runs reflects ensemble variability across independently generated networks and, when results are pooled, across the parameter sweep, rather than intrinsic stochasticity in transmission.

\begin{figure*}[t]
\centering
    \includegraphics[width=0.98\textwidth]{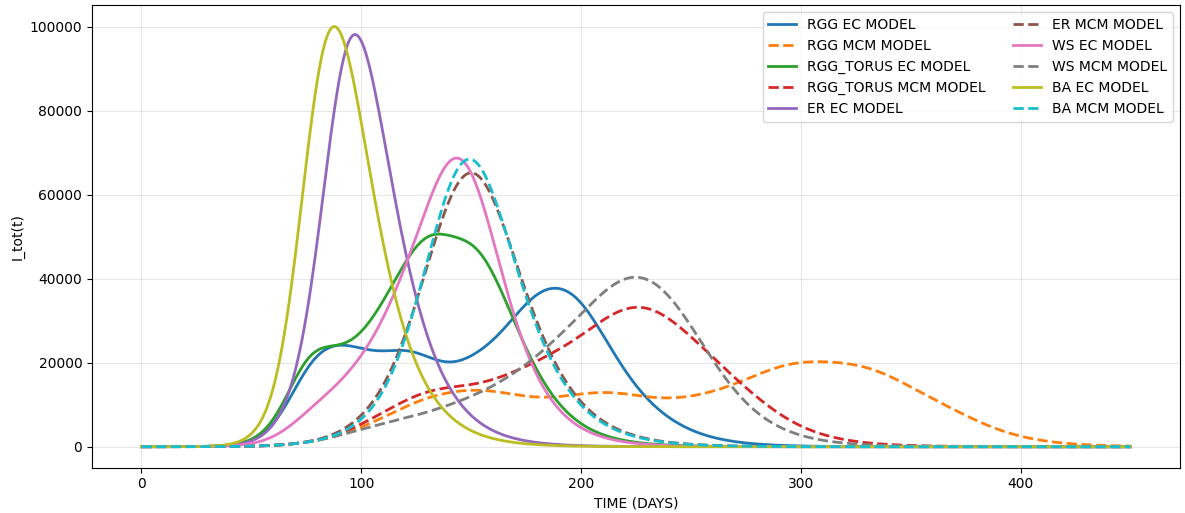}
\caption{
Representative aggregate epidemic trajectories \(I_{\mathrm{tot}}(t)\) for the EC and MCM coupling formulations across the different network topologies.
Solid curves correspond to the effective-coupling model (EC), while dashed curves correspond to the mobility-consistent mixing model (MCM).
The comparison includes the square RGG, a degree-matched toroidal RGG, and degree-matched non-spatial ER, WS, and BA networks.
The toroidal RGG radius was chosen to match, in expectation, the realized
mean degree of the corresponding square RGG.
The spatially constrained RGGs exhibit delayed and broader epidemic trajectories, including visible multi-peak structure, whereas the non-spatial degree-matched networks tend to synchronize more rapidly and produce sharper aggregate peaks.
These trajectories correspond to the same representative network realizations used in the spatiotemporal heatmap visualizations shown in Figs.~\ref{fig:heatmaps-ec} and~\ref{fig:heatmaps-mcm}, providing the aggregate temporal context for the node-level propagation patterns displayed there.}
\label{fig:ec_mcm_combined}
\end{figure*}

\begin{table}[htbp]
\caption{Network parameters for the representative heatmap configuration.}
\label{tab:network_parameters_heatmap}

\setlength{\tabcolsep}{7pt}
\begin{tabular}{lcccc}
\hline
Topology & Param. & \(n\) & Edges \(|E|\) & \(\langle k \rangle\) \\
\hline
RGG & \(r=0.110\) & 200 & 628 & 6.28 \\
RGG torus & \(r_{\mathrm{torus}}=0.10023\) & 200 & 649 & 6.49 \\
ER & \(p=0.0316\) & 200 & 613 & 6.13 \\
BA & \(m=3\) & 200 & 591 & 5.91 \\
WS & \(k=6,\, p=0.050\) & 200 & 600 & 6.00 \\
\hline
\end{tabular}
\end{table}

We compare network topologies separately under EC and MCM dynamics.
This allows us to ask whether the effect of RGG-like spatial constraints is specific to one coupling formulation or persists across different assumptions about inter-node transmission.
In both formulations, square RGGs show delayed spatial invasion and stronger local outbreak asynchrony than degree-matched ER, WS, and BA baselines.
The toroidal RGG reduces boundary-driven delays while preserving local geometric connectivity, producing intermediate behavior between the square RGG and the non-spatial baselines.
Thus, the qualitative RGG effect is robust across coupling formulations, although its magnitude differs between EC and MCM.

\subsection{Aggregate trajectories and multi-peak behavior}
\label{subsec:results_trajectories}

A central dynamical outcome of interest is the emergence of multiple epidemic waves in the aggregate infected trajectory $I_{\mathrm{tot}}(t).$
For each simulation, multi-peak trajectories were identified using the fixed peak-detection protocol described in Section~\ref{subsec:peaks}.
A trajectory was classified as multi-peak when at least two peaks satisfying the prescribed prominence and separation criteria were detected.

\begin{figure*}[t]
  \centering
  \includegraphics[width=0.98\textwidth]{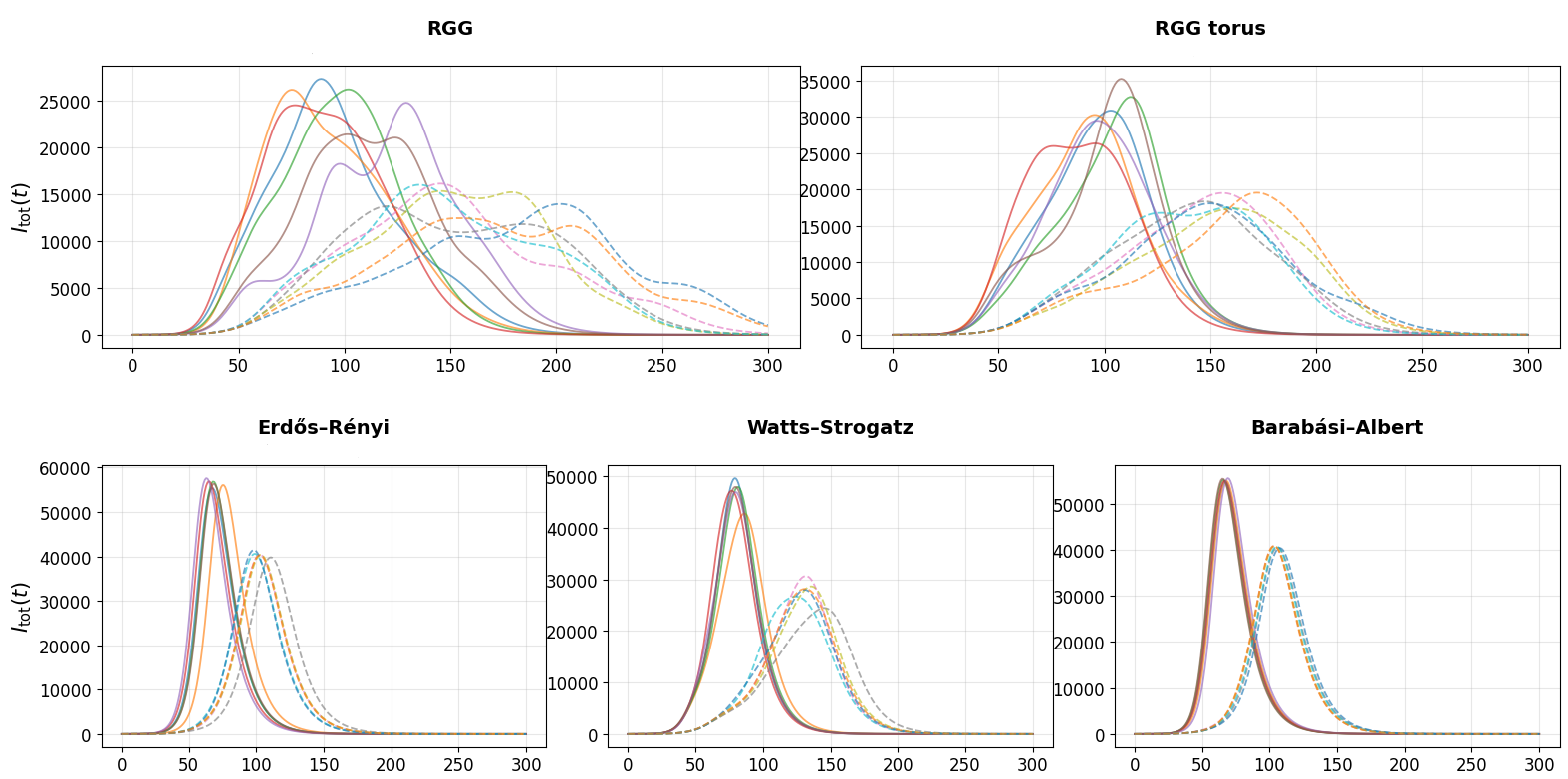}
  \caption{
Representative individual aggregate epidemic trajectories selected from the simulation ensembles for different network topologies.
The comparison includes the square RGG, the degree-matched toroidal RGG, and the degree-matched non-spatial Erd\H{o}s--R\'enyi, Watts--Strogatz, and Barab\'asi--Albert baselines.
For each topology, six realizations are shown to illustrate the variability of trajectory shapes within the ensemble.
Solid curves correspond to the effective-coupling model (EC), whereas dashed curves correspond to the mobility-consistent mixing model (MCM).
The spatially constrained RGGs display broader, delayed, and more visibly multi-peak aggregate trajectories, while the non-spatial degree-matched networks tend to produce more synchronized epidemic peaks.
Quantitative comparisons are based on ensemble-level summaries reported below. $n=200,r=0.11, N_i=1000, \beta=0.25,\gamma=0.1,\delta=0.03.$
}
  \label{fig:representative_trajectories}
\end{figure*}

Figure~\ref{fig:representative_trajectories} illustrates the main qualitative difference between spatially constrained and non-spatial degree-matched networks.
RGG trajectories are broader and more irregular, with delayed shoulders or multiple peaks in several realizations.
In contrast, ER, WS, and BA baselines tend to produce more synchronized aggregate outbreaks.
The toroidal RGG shows intermediate behavior: removing hard boundaries reduces the multi-peak tendency, but does not make the dynamics identical to the non-spatial baselines.

Tables~\ref{tab:multipeak_ec_compact_with_torus} and~\ref{tab:multipeak_mcm_compact_with_torus} report the multi-peak percentages across network topologies and RGG configurations for EC and MCM dynamics, respectively.
Across the explored configurations, square RGGs consistently exhibit higher multi-peak frequencies than the degree-matched non-spatial baselines.
Toroidal RGGs reduce the frequency of multi-peak trajectories relative to square RGGs, indicating that hard boundaries contribute to the effect.
However, multi-peak behavior remains present in several toroidal configurations, showing that the phenomenon is not solely a boundary artifact.

\begin{table*}[!htbp]
\centering
\footnotesize
\setlength{\tabcolsep}{10.5pt}
\renewcommand{\arraystretch}{1.25}

\begin{minipage}[t]{0.485\textwidth}
\centering
\captionof{table}{Multi-peak outcomes for the EC model across topologies and configurations.}
\label{tab:multipeak_ec_compact_with_torus}
\begin{tabular}{cccc}
\toprule
$n$ & $r$ & Topology & M-peaks (\%) \\
\midrule
200 & 0.095 & RGG  & 60.0 \\
200 & 0.095 & RGG$_{\mathrm{torus}}$ & 29.1 \\
200 & 0.095 & ER & 0.0 \\
200 & 0.095 & WS & 0.0 \\
200 & 0.095 & BA & 0.0 \\
\midrule
200 & 0.110 & RGG & 28.7 \\
200 & 0.110 & RGG$_{\mathrm{torus}}$ & 0.6 \\
200 & 0.110 & ER & 0.0 \\
200 & 0.110 & WS & 0.0 \\
200 & 0.110 & BA & 0.0 \\
\midrule
200 & 0.125 & RGG & 7.2 \\
200 & 0.125 & RGG$_{\mathrm{torus}}$ & 0.0 \\
200 & 0.125 & ER & 0.0 \\
200 & 0.125 & WS & 0.0 \\
200 & 0.125 & BA & 0.0 \\
\midrule
400 & 0.075 & RGG & 43.8 \\
400 & 0.075 & RGG$_{\mathrm{torus}}$ & 5.3 \\
400 & 0.075 & ER & 0.0 \\
400 & 0.075 & WS & 0.0 \\
400 & 0.075 & BA & 0.0 \\
\midrule
400 & 0.090 & RGG & 13.1 \\
400 & 0.090 & RGG$_{\mathrm{torus}}$ & 0.9 \\
400 & 0.090 & ER & 0.0 \\
400 & 0.090 & WS & 0.0 \\
400 & 0.090 & BA & 0.0 \\
\midrule
400 & 0.105 & RGG & 1.9 \\
400 & 0.105 & RGG$_{\mathrm{torus}}$ & 0.0 \\
400 & 0.105 & ER & 0.0 \\
400 & 0.105 & WS & 0.0 \\
400 & 0.105 & BA & 0.0 \\
\bottomrule
\end{tabular}
\end{minipage}
\hfill
\vrule width 0.4pt
\hfill
\begin{minipage}[t]{0.485\textwidth}
\centering
\captionof{table}{Multi-peak outcomes for the MCM model across topologies and configurations.}
\label{tab:multipeak_mcm_compact_with_torus}
\begin{tabular}{cccc}
\toprule
$n$ & $r$ & Topology & M-peaks (\%) \\
\midrule
200 & 0.095 & RGG & 56.6 \\
200 & 0.095 & RGG$_{\mathrm{torus}}$ & 29.1 \\
200 & 0.095 & ER & 0.0 \\
200 & 0.095 & WS & 0.0 \\
200 & 0.095 & BA & 0.0 \\
\midrule
200 & 0.110 & RGG & 37.5 \\
200 & 0.110 & RGG$_{\mathrm{torus}}$ & 2.8 \\
200 & 0.110 & ER & 0.0 \\
200 & 0.110 & WS & 0.0 \\
200 & 0.110 & BA & 0.0 \\
\midrule
200 & 0.125 & RGG & 22.2 \\
200 & 0.125 & RGG$_{\mathrm{torus}}$ & 0.0 \\
200 & 0.125 & ER & 0.0 \\
200 & 0.125 & WS & 0.0 \\
200 & 0.125 & BA & 0.0 \\
\midrule
400 & 0.075 & RGG & 45.0 \\
400 & 0.075 & RGG$_{\mathrm{torus}}$ & 4.7 \\
400 & 0.075 & ER & 0.0 \\
400 & 0.075 & WS & 0.0 \\
400 & 0.075 & BA & 0.0 \\
\midrule
400 & 0.090 & RGG & 26.9 \\
400 & 0.090 & RGG$_{\mathrm{torus}}$ & 0.3 \\
400 & 0.090 & ER & 0.0 \\
400 & 0.090 & WS & 0.0 \\
400 & 0.090 & BA & 0.0 \\
\midrule
400 & 0.105 & RGG & 15.0 \\
400 & 0.105 & RGG$_{\mathrm{torus}}$ & 0.0 \\
400 & 0.105 & ER & 0.0 \\
400 & 0.105 & WS & 0.0 \\
400 & 0.105 & BA & 0.0 \\
\bottomrule
\end{tabular}
\end{minipage}

\vspace{0.15cm}
\begin{minipage}{0.95\textwidth}
\footnotesize
\textit{Note.}
Here, \(n\) is the number of network nodes, and \(r\) is the connection
radius of the reference square RGG configuration. The corresponding
toroidal RGG and non-spatial baseline networks were generated using the
degree-matching procedure described in Section~\ref{subsec:network_ensembles}.
M-peaks gives the percentage of simulations classified as multi-peak
according to the fixed peak-detection protocol described in
Section~\ref{subsec:peaks}.
\end{minipage}

\end{table*}

\subsection{Spatiotemporal heatmaps reveal delayed spatial invasion}
\label{subsec:results_heatmaps}
To visualize the node-level organization underlying the aggregate trajectories, we
generated heatmap snapshots of local infected counts \(I_i(t)\) at selected times.
These heatmaps illustrate how infection propagates across the network and how local
outbreaks become synchronized or desynchronized across topologies.
The aggregate epidemic trajectories corresponding to these same representative
network realizations are shown in Fig.~\ref{fig:ec_mcm_combined}, providing the temporal context
for the node-level propagation patterns displayed below.

\begin{figure*}[p]
\centering

\begingroup
\setlength{\tabcolsep}{0pt}

\makebox[\textwidth][c]{%
\begin{tabular}{cccc}

\textbf{$t_1$} & \textbf{$t_2$} & \textbf{$t_3$} & \textbf{$t_4$} \\[2mm]

\multicolumn{4}{c}{\textbf{Square RGG}} \\[1mm]
\includegraphics[width=0.225\textwidth]{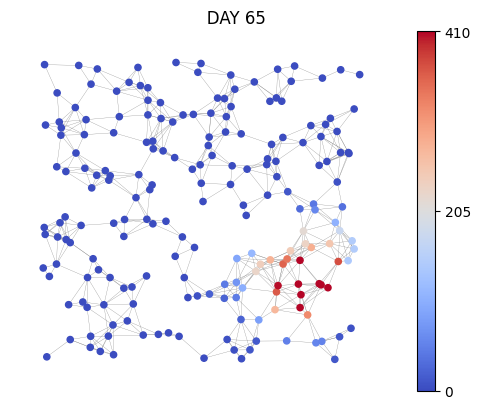} &
\includegraphics[width=0.225\textwidth]{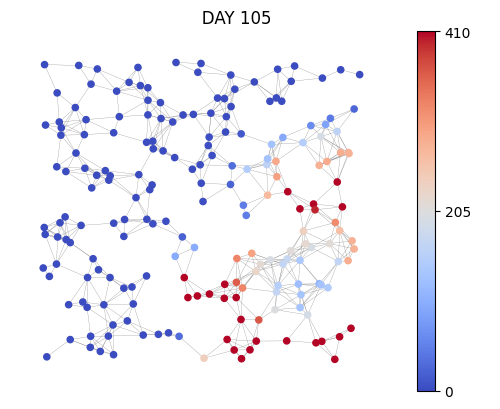} &
\includegraphics[width=0.225\textwidth]{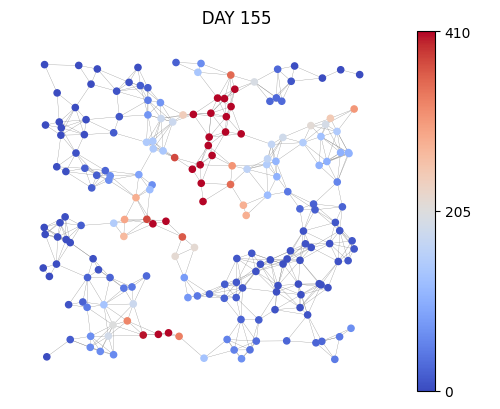} &
\includegraphics[width=0.225\textwidth]{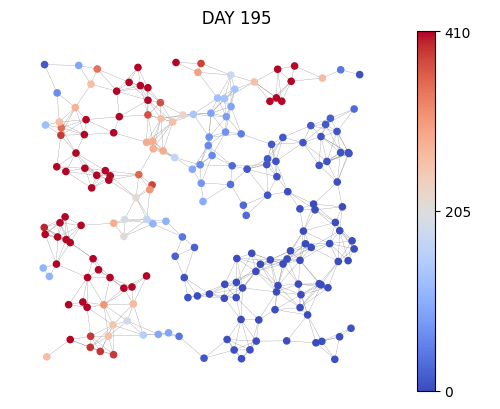} \\[1mm]

\multicolumn{4}{c}{\textbf{Toroidal RGG}} \\[1mm]
\includegraphics[width=0.225\textwidth]{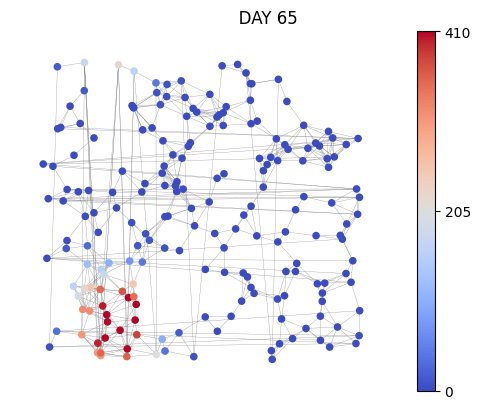} &
\includegraphics[width=0.225\textwidth]{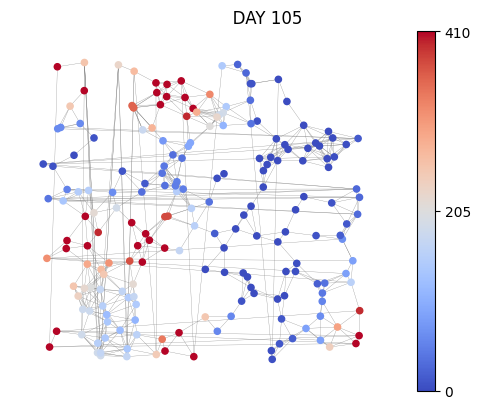} &
\includegraphics[width=0.225\textwidth]{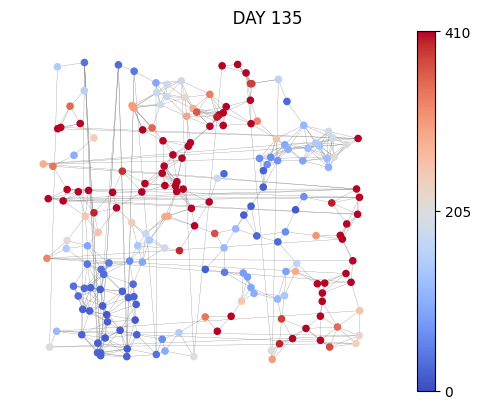} &
\includegraphics[width=0.225\textwidth]{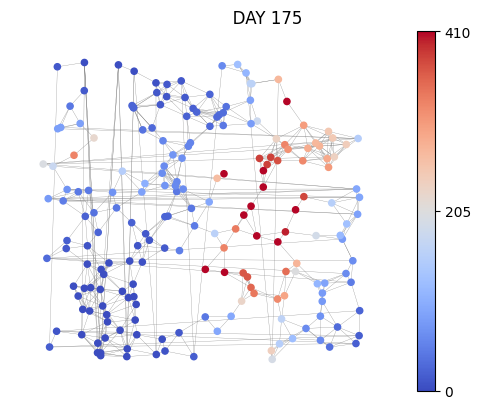} \\[1mm]

\multicolumn{4}{c}{\textbf{Erd\H{o}s--R\'enyi (ER)}} \\[1mm]
\includegraphics[width=0.225\textwidth]{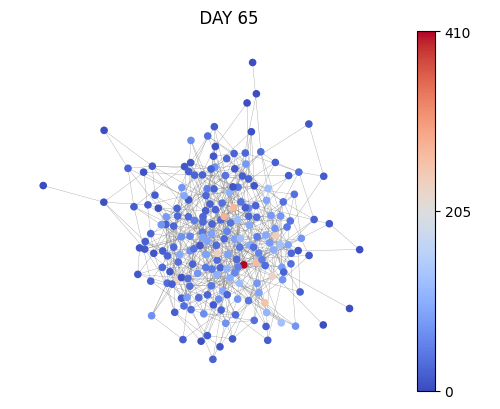} &
\includegraphics[width=0.225\textwidth]{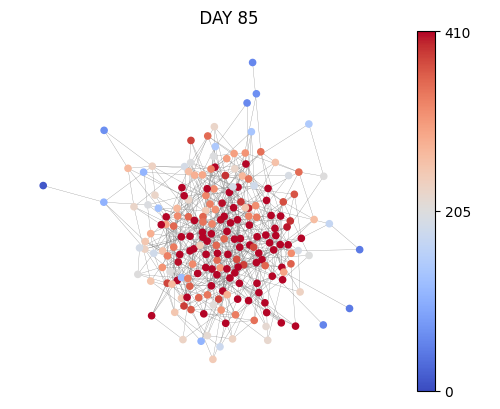} &
\includegraphics[width=0.225\textwidth]{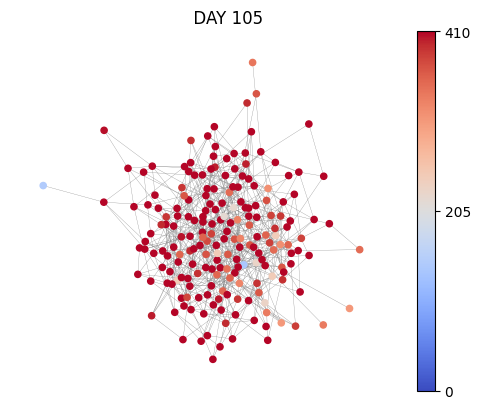} &
\includegraphics[width=0.225\textwidth]{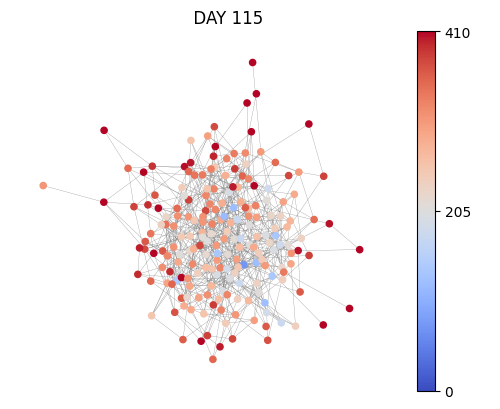} \\[1mm]

\multicolumn{4}{c}{\textbf{Watts--Strogatz (WS)}} \\[1mm]
\includegraphics[width=0.225\textwidth]{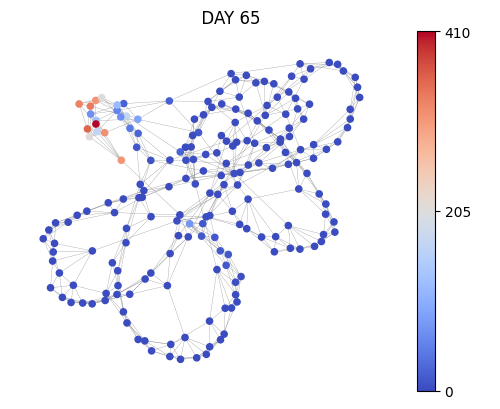} &
\includegraphics[width=0.225\textwidth]{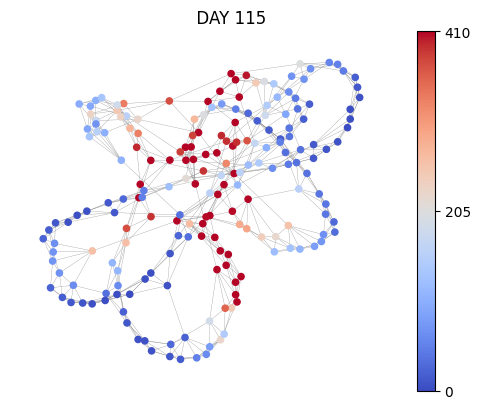} &
\includegraphics[width=0.225\textwidth]{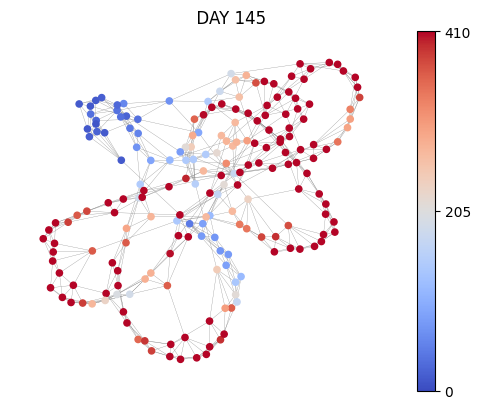} &
\includegraphics[width=0.225\textwidth]{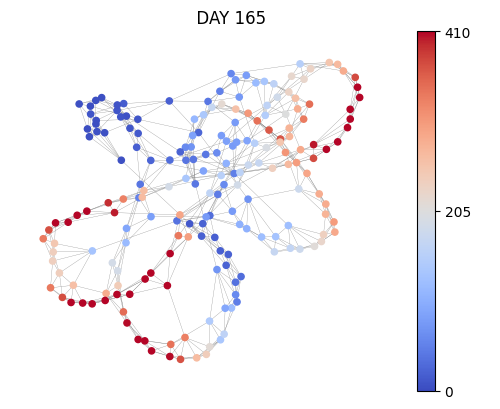} \\[1mm]

\multicolumn{4}{c}{\textbf{Barab\'asi--Albert (BA)}} \\[1mm]
\includegraphics[width=0.225\textwidth]{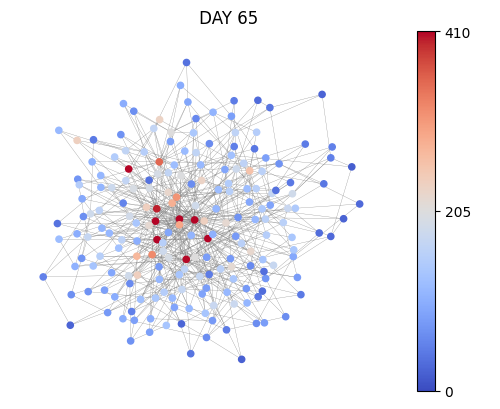} &
\includegraphics[width=0.225\textwidth]{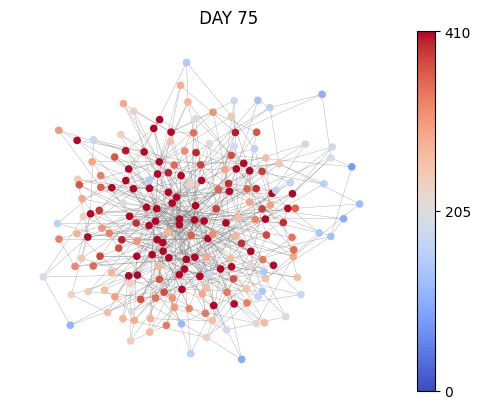} &
\includegraphics[width=0.225\textwidth]{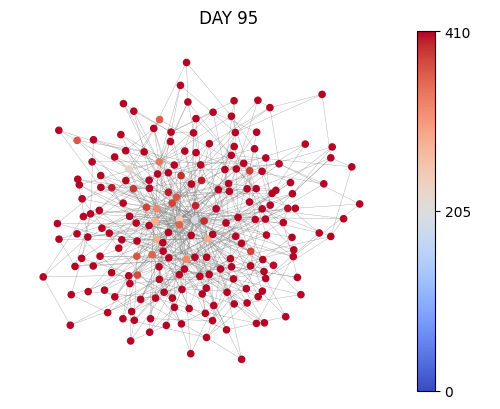} &
\includegraphics[width=0.225\textwidth]{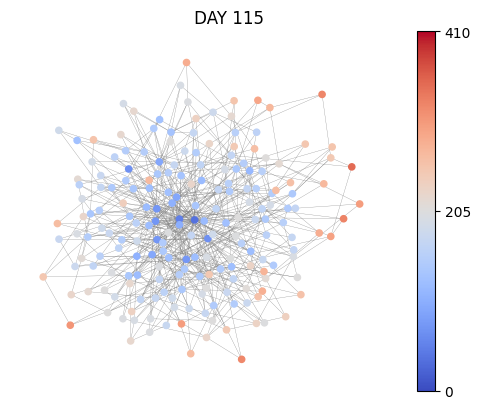} \\

\end{tabular}%
}

\endgroup

\caption{Spatiotemporal heatmap snapshots for the effective-coupling model
(EC). Columns \(t_1,\ldots,t_4\) denote topology-specific representative
stages of epidemic evolution rather than common absolute times across rows.
Node colors indicate the number of infected individuals \(I_i(t)\).
All panels use the fixed color scale \(v_{\min}=0\) and \(v_{\max}=410\).}
\label{fig:heatmaps-ec}
\end{figure*}

\begin{figure*}[p]
\centering

\begingroup
\setlength{\tabcolsep}{0pt}

\makebox[\textwidth][c]{%
\begin{tabular}{cccc}

\textbf{$t_1$} & \textbf{$t_2$} & \textbf{$t_3$} & \textbf{$t_4$} \\[2mm]

\multicolumn{4}{c}{\textbf{Square RGG}} \\[1mm]
\includegraphics[width=0.225\textwidth]{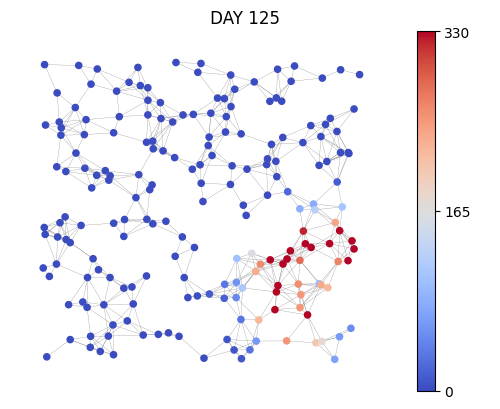} &
\includegraphics[width=0.225\textwidth]{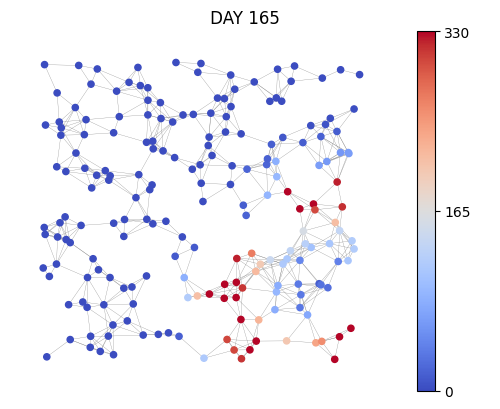} &
\includegraphics[width=0.225\textwidth]{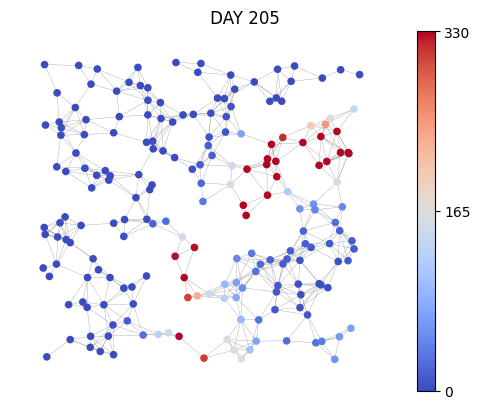} &
\includegraphics[width=0.225\textwidth]{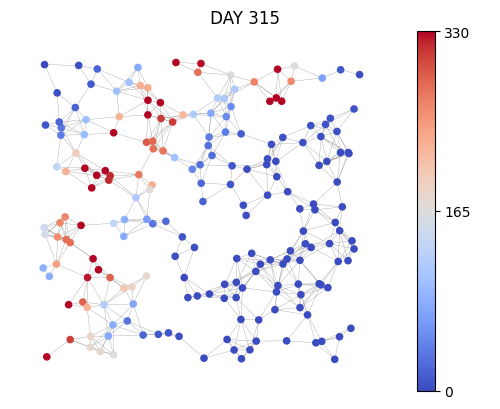} \\[1mm]

\multicolumn{4}{c}{\textbf{Toroidal RGG}} \\[1mm]
\includegraphics[width=0.225\textwidth]{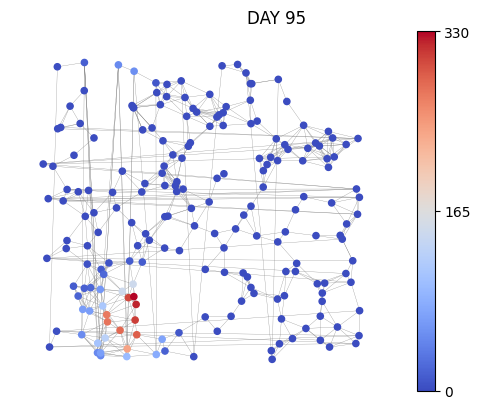} &
\includegraphics[width=0.225\textwidth]{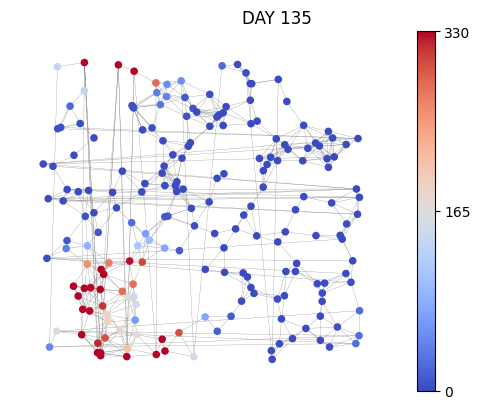} &
\includegraphics[width=0.225\textwidth]{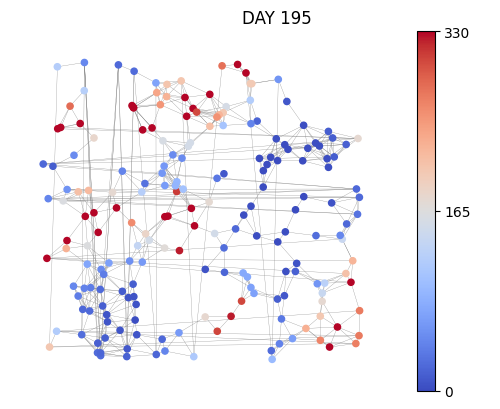} &
\includegraphics[width=0.225\textwidth]{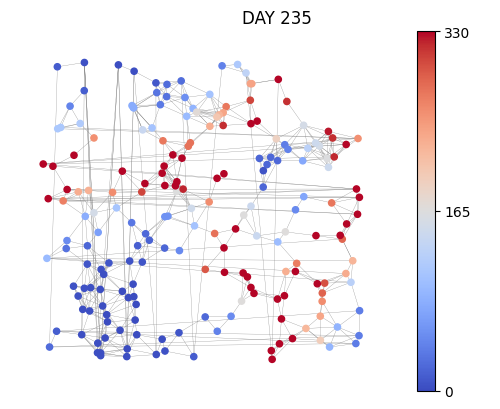} \\[1mm]

\multicolumn{4}{c}{\textbf{Erd\H{o}s--R\'enyi (ER)}} \\[1mm]
\includegraphics[width=0.225\textwidth]{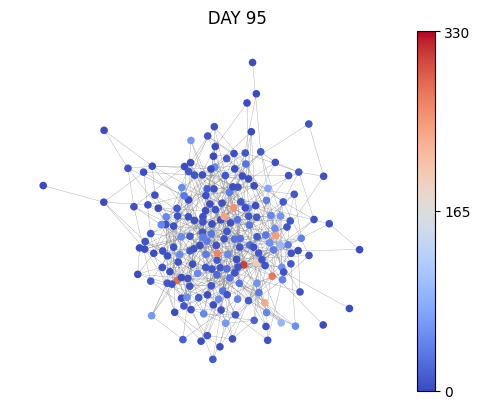} &
\includegraphics[width=0.225\textwidth]{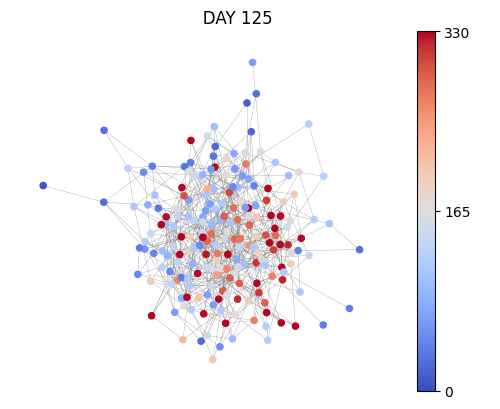} &
\includegraphics[width=0.225\textwidth]{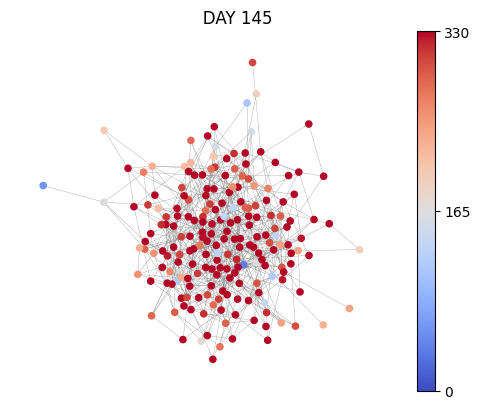} &
\includegraphics[width=0.225\textwidth]{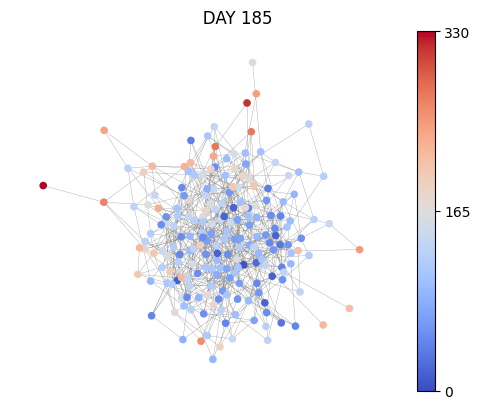} \\[1mm]

\multicolumn{4}{c}{\textbf{Watts--Strogatz (WS)}} \\[1mm]
\includegraphics[width=0.225\textwidth]{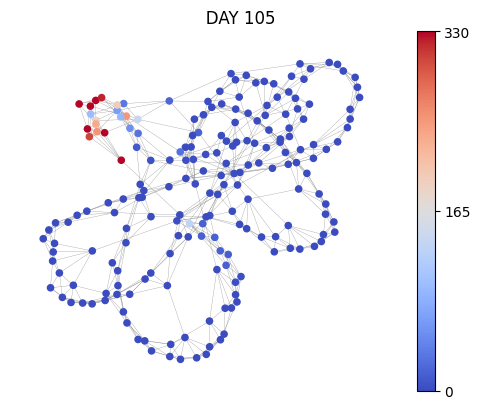} &
\includegraphics[width=0.225\textwidth]{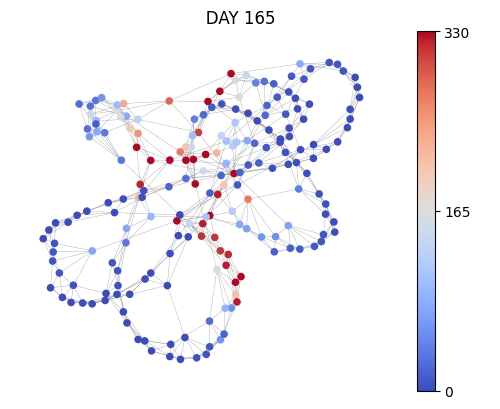} &
\includegraphics[width=0.225\textwidth]{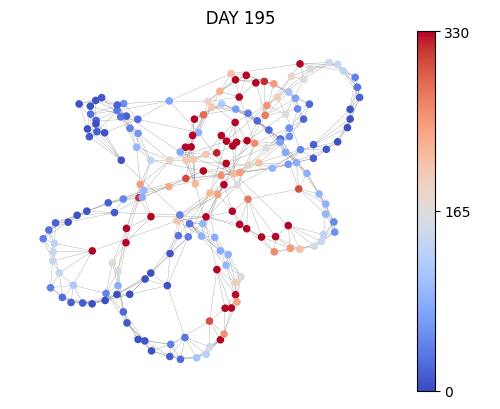} &
\includegraphics[width=0.225\textwidth]{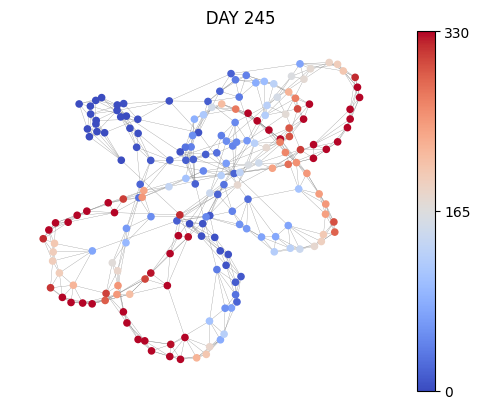} \\[1mm]

\multicolumn{4}{c}{\textbf{Barab\'asi--Albert (BA)}} \\[1mm]
\includegraphics[width=0.225\textwidth]{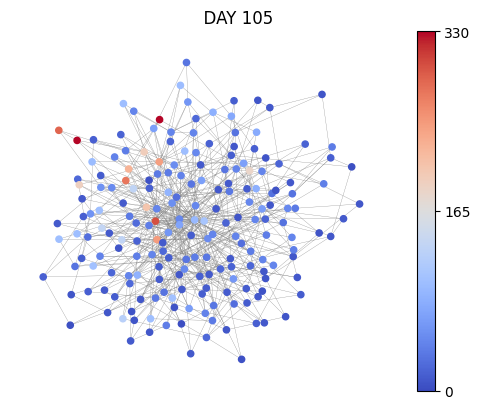} &
\includegraphics[width=0.225\textwidth]{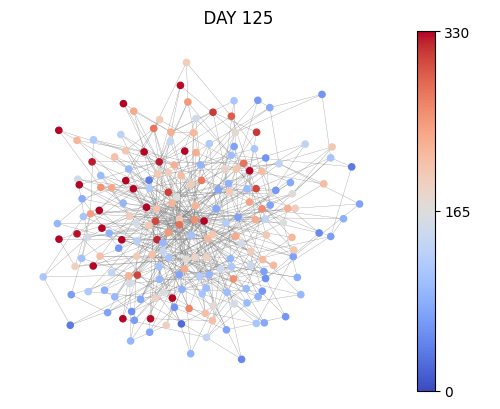} &
\includegraphics[width=0.225\textwidth]{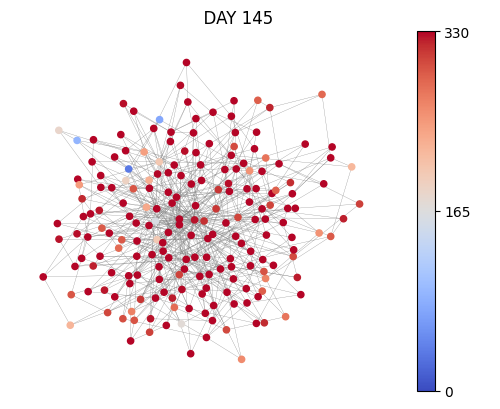} &
\includegraphics[width=0.225\textwidth]{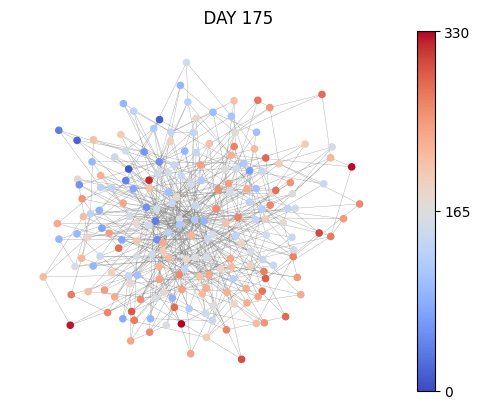} \\

\end{tabular}%
}

\endgroup

\caption{Spatiotemporal heatmap snapshots for the mobility-consistent mixing
model (MCM). Columns \(t_1,\ldots,t_4\) denote topology-specific
representative stages of epidemic evolution rather than common absolute
times across rows. Node colors indicate the number of infected individuals
\(I_i(t)\). All panels use the fixed color scale
\(v_{\min}=0\) and \(v_{\max}=319\).}
\label{fig:heatmaps-mcm}
\end{figure*}

The heatmaps show the local mechanism behind the aggregate trajectory differences.
In the square RGG, infection remains spatially localized for longer periods and spreads through separated regions of the network, producing delayed activation of distant nodes.
The toroidal RGG reduces this delay but still preserves local geometric propagation.
In contrast, ER and BA networks activate much more synchronously, and WS networks show intermediate behavior depending on the extent to which local neighborhood structure is preserved.
These visual patterns are consistent with the higher multi-peak frequencies
observed for spatially constrained RGGs across the parameter sweep.

\section*{Discussion}

This work examined how local geometric connectivity affects epidemic propagation in networked metapopulation systems.
Using random geometric graphs (RGGs) as controlled spatial network ensembles, we showed that spatially constrained connectivity can delay epidemic invasion, desynchronize local outbreaks, and generate broad, shouldered, or multi-peak aggregate epidemic trajectories.
The central mechanism is not a change in local epidemic dynamics, but a change in the temporal organization of spatial spread: infection must propagate through local graph neighborhoods, producing staggered local outbreaks whose superposition shapes the aggregate infected curve.

\paragraph{Spatial constraints promote delayed invasion and multi-peak dynamics.}
Multi-peak trajectories were consistently observed in the square RGG
ensembles, were reduced but remained present in some toroidal RGG
configurations, and were absent from all degree-matched ER, WS, and BA
baselines across the explored parameter sweep.
This pattern shows that differences in mean connectivity alone are insufficient
to account for the observed trajectory shapes.
The representative trajectories and spatiotemporal visualizations further show that
RGGs sustain more localized and temporally staggered epidemic propagation,
so that local outbreaks contribute asynchronously to the aggregate infected
curve.
When this temporal separation is sufficiently pronounced, the aggregate
signal can exhibit shoulders, extended plateaus, or multiple peaks.

\paragraph{Boundary effects and toroidal RGGs.}
The toroidal RGG control helps distinguish the contribution of hard boundaries
from the broader effect of local geometric connectivity.
Removing hard boundaries substantially reduces the frequency of multi-peak
trajectories relative to square RGGs, but does not eliminate them.
Thus, hard boundaries amplify the observed behavior but do not fully account
for it.
The persistence of multi-peak trajectories under periodic boundary conditions
supports the interpretation that local geometric connectivity itself
contributes to delayed and asynchronous epidemic propagation.

\paragraph{Role of the coupling formulations.}
The EC and MCM formulations were used as mechanistic alternatives to test whether the spatial-network effects depend on a particular representation of inter-node transmission.
Under the homogeneous assumptions considered here, the MCM formulation yields the threshold relation \(R_0=\beta/\gamma\), independently of the adjacency structure, whereas in the EC formulation the basic reproduction number depends on the spectral properties of the network coupling matrix.
This distinction is useful for interpreting the results.
In particular, delayed invasion, local outbreak asynchrony, and multi-peak aggregate trajectories are also observed in spatially constrained RGGs under MCM, even though topology does not modify \(R_0\) in that formulation.
Therefore, the observed topology-dependent differences cannot be explained solely by changes in the epidemic threshold or basic reproduction number.
Instead, these results support the interpretation that network geometry shapes the transient spatial propagation and timing of local outbreaks.
The magnitude of these effects differs between EC and MCM because the two formulations encode different assumptions about inter-node transmission, but the qualitative RGG-driven mechanism persists under both.
Thus, the coupling comparison should be interpreted as a robustness check rather than as the central object of the study.

\section*{Limitations}\label{sec:limitations}

The present study is intentionally focused on a stylized metapopulation
framework, and several limitations should be acknowledged when interpreting
the results.

First, the spatial networks considered here are controlled network ensembles
rather than empirical reconstructions of a particular transportation, contact,
or mobility system. Random geometric graphs provide a minimal model of
spatially embedded connectivity: they capture locality, geometric constraints,
neighborhood overlap, and high clustering, but they do not encode real-world
geography, population centers, infrastructure, or long-range travel.
Consequently, RGGs should be interpreted as a controlled testbed for studying
the effects of local geometric connectivity rather than as data-driven
representations of specific regions or mobility systems. Similarly, the
toroidal RGG is used as a structural control for boundary effects, not as a
realistic geographical model. The present analysis is therefore intended as a
mechanistic study rather than as a data-driven representation of a specific
epidemic system.

Second, changing the connection radius in an RGG modifies several structural
properties simultaneously, including mean degree, clustering, connectivity,
and other aspects of network organization. Although the degree-matched
baselines control for large differences in average connectivity, the present
study does not attempt to isolate the independent causal contribution of every
structural network descriptor. Our interpretation is therefore mechanistic
rather than exhaustive: the simulations examine the collective effect of local
geometric connectivity on epidemic propagation rather than attributing the
observed dynamics to a single graph metric.

Third, the coupling formulations used in this work are idealized
representations of inter-node transmission. The mobility-consistent mixing
model (MCM) represents a simplified local mixing-at-destination mechanism, in
which individuals distribute mobility only among neighboring subpopulations.
It does not include long-range travel, directional commuting patterns,
heterogeneous residence times, or empirically measured mobility flows. The
effective-coupling model (EC), in turn, represents edge-mediated infection
pressure without explicitly modeling population movement. Thus, EC and MCM
should be viewed as mechanistic alternatives used to test robustness across
coupling assumptions, rather than as calibrated descriptions of a specific
epidemic system.

Fourth, the epidemiological dynamics are deterministic and homogeneous within
each subpopulation. Unless otherwise stated, all nodes have the same local
population size and share the same transmission and recovery parameters.
Population heterogeneity, behavioral changes, interventions, seasonality,
stochastic transmission, pathogen evolution, and reporting effects can
introduce additional temporal and spatial variability that is not represented
here.

Finally, the results should not be interpreted as implying that multi-peak
epidemic curves arise only from spatial network structure. Seasonality,
interventions, behavioral responses, parameter heterogeneity, stochastic
effects, and data aggregation can also generate complex aggregate epidemic
profiles. The more specific conclusion of this study is that local geometric
connectivity provides one structural mechanism through which simple local
epidemic dynamics can produce delayed invasion, local outbreak asynchrony,
and broad or multi-peak aggregate trajectories.

\section*{Concluding remarks}

Overall, our results support local geometric connectivity as a structural
mechanism linking network geometry to delayed invasion, local outbreak
asynchrony, and complex aggregate epidemic trajectories. Spatially constrained
RGGs generated broad and multi-peak aggregate dynamics despite simple and
homogeneous local epidemic dynamics, whereas multi-peak trajectories were
absent from all degree-matched ER, WS, and BA baselines across the explored
parameter sweep. This shows that mean connectivity alone is insufficient to
account for the observed differences in epidemic trajectory shape.

The toroidal RGG control further shows that hard boundaries amplify the
phenomenon but do not fully account for it: multi-peak behavior persists under
periodic boundary conditions, although at substantially lower frequencies than
in square RGGs. Moreover, the persistence of delayed invasion, local outbreak asynchrony,
and multi-peak dynamics under the MCM formulation, for which
\(R_0=\beta/\gamma\) is independent of network topology under the
homogeneous assumptions considered here, shows that these topology-dependent
differences cannot be attributed solely to changes in the epidemic threshold.
Instead, they support the interpretation that network geometry shapes the
transient spatial propagation and relative timing of local outbreaks.

Together, these results show that epidemic trajectory shape depends not only
on mean connectivity, but also on how connections are organized in space.
By staggering the timing of otherwise simple local outbreaks, local geometric
constraints can produce an asynchronous superposition of epidemic activity
that gives rise to broad, shouldered, or multi-peak aggregate trajectories.

\section*{Data availability}
The data supporting the findings of this study consist of simulation outputs
 generated using the computational framework described in the manuscript.
The source code and processed simulation datasets are available from the
corresponding author upon reasonable request.

\begin{acknowledgments}
This work was supported by the Research Group Project (PGI-24/L131) of the Universidad Nacional del Sur (Argentina).
\end{acknowledgments}

\bibliography{referencias_bibtex2}
\end{document}